\documentclass[prx,reprint,superscriptaddress,longbibliography]{revtex4-2}

\usepackage{mathtools}
\usepackage{hyperref}
\usepackage{multirow}
\usepackage{placeins}  % to use \FloatBarrier

\usepackage{mleftright}
\mleftright  % to fix spacing issues with \left & \right

\newcommand{\mr}[1]{\mathrm{#1}}
\newcommand{\mc}[1]{\mathcal{#1}}
\newcommand{\hmc}[1]{\hat{\mathcal{#1}}}
\newcommand{\hc}{\hat{c}}
\newcommand{\hcd}{\hat{c}^\dag}
\newcommand{\hf}{\hat{f}}
\newcommand{\hfd}{\hat{f}^\dag}
\newcommand{\hb}{\hat{b}}
\newcommand{\hbd}{\hat{b}^\dag}

\DeclarePairedDelimiter{\ket}{\lvert}{\rangle}
\DeclarePairedDelimiter{\expect}{\langle}{\rangle}
\DeclarePairedDelimiterX{\braket}[2]{\langle}{\rangle}{#1\delimsize\vert\mathopen{}#2}
\DeclarePairedDelimiterX{\ketbra}[2]{\lvert}{\rvert}{#1\delimsize\rangle\!\delimsize\langle#2}

\newcommand{\overbar}[1]{\mkern 1.5mu\overline{\mkern-1.5mu#1\mkern-1.5mu}\mkern 1.5mu}

\begin{document}

%TC:ignore

\title{Photon-noise-tolerant dispersive readout of a superconducting qubit\\ using a nonlinear Purcell filter}

\author{Yoshiki~Sunada}
\email{yoshiki.sunada@aalto.fi}
\altaffiliation[Present address: ]{QCD Labs, QTF Centre of Excellence, Department of Applied Physics, Aalto University, P.O.\ Box 13500, FIN-00076 Aalto, Finland}
\affiliation{Department of Applied Physics, Graduate School of Engineering, The University of Tokyo, Bunkyo-ku, Tokyo 113-8656, Japan}
\author{Kenshi~Yuki}
\affiliation{Department of Applied Physics, Graduate School of Engineering, The University of Tokyo, Bunkyo-ku, Tokyo 113-8656, Japan}
\author{Zhiling~Wang}
\affiliation{RIKEN Center for Quantum Computing (RQC), Wako, Saitama 351-0198, Japan}
\author{Takeaki~Miyamura}
\affiliation{Department of Applied Physics, Graduate School of Engineering, The University of Tokyo, Bunkyo-ku, Tokyo 113-8656, Japan}
\author{Jesper~Ilves}
\affiliation{Department of Applied Physics, Graduate School of Engineering, The University of Tokyo, Bunkyo-ku, Tokyo 113-8656, Japan}
\author{Kohei~Matsuura}
\affiliation{Department of Applied Physics, Graduate School of Engineering, The University of Tokyo, Bunkyo-ku, Tokyo 113-8656, Japan}
\author{Peter~A.~Spring}
\affiliation{RIKEN Center for Quantum Computing (RQC), Wako, Saitama 351-0198, Japan}
\author{Shuhei~Tamate}
\affiliation{RIKEN Center for Quantum Computing (RQC), Wako, Saitama 351-0198, Japan}
\author{Shingo~Kono}
\affiliation{Institute of Physics, Swiss Federal Institute of Technology Lausanne (EPFL), Lausanne, Switzerland}
\affiliation{Center for Quantum Science and Engineering, EPFL, Lausanne, Switzerland}
\author{Yasunobu~Nakamura}
\affiliation{Department of Applied Physics, Graduate School of Engineering, The University of Tokyo, Bunkyo-ku, Tokyo 113-8656, Japan}
\affiliation{RIKEN Center for Quantum Computing (RQC), Wako, Saitama 351-0198, Japan}

\date{\today}

\begin{abstract}

Residual noise photons in a readout resonator become a major source of dephasing for a superconducting qubit when the resonator is optimized for a fast, high-fidelity dispersive readout.
Here, we propose and demonstrate a nonlinear Purcell filter that suppresses such an undesirable dephasing process without sacrificing the readout performance.
When a readout pulse is applied, the filter automatically reduces the effective linewidth of the readout resonator, increasing the sensitivity of the qubit to the input field.
The noise tolerance of the device we have fabricated is shown to be enhanced by a factor of 3 relative to a device with a linear filter.
The measurement rate is enhanced by another factor of 3 by utilizing the bifurcation of the nonlinear filter.
A readout fidelity of 99.4\% and a quantum nondemolition fidelity of 99.2\% are achieved using a 40-ns readout pulse.
The nonlinear Purcell filter will be an effective tool for realizing a fast, high-fidelity readout without compromising the coherence time of the qubit.

\end{abstract}

\maketitle

%TC:endignore

\section{INTRODUCTION}

Dispersive readout has been remarkably successful as a technique to quickly and accurately measure the state of a superconducting qubit~\cite{blais200406cavity,walter201705rapid}.
It is based on the qubit-state-dependent frequency shift of a resonator dispersively coupled to the qubit.
However, the resonator introduces additional decoherence channels to the qubit.
One such channel is the energy relaxation of the qubit through the resonator, which can be suppressed using a ``Purcell filter''~\cite{reed201005fast,jeffrey201405fast, sete201507quantum,bronn201510reducing,bronn201510broadband,govia201702enhanced,sunada202204fast,yan202306broadband} or a large qubit--resonator detuning~\cite{nguyen202208blueprint}.
Another prominent decoherence process is the dephasing due to residual noise photons in the resonator~\cite{bertet200512dephasing,clerk200704using,serban200710crossover,rigetti201209superconducting,sears201211photon,yan201611flux,yan201806distinguishing}.
The conventional countermeasure against the photon-noise-induced dephasing has been to suppress the influx of noise from the readout waveguide~\cite{yeh201706microwave,wang201901cavity,abdo202112highfidelity}.
A complementary approach would be to mitigate the adverse effect of the noise on the device side, but, to our knowledge, no such strategy has been proposed.

Here, we propose and demonstrate a nonlinear Purcell filter that suppresses the photon-noise-induced dephasing of the qubit without sacrificing the readout performance.
The filter is nonlinear in the sense that its transmission coefficient depends on the amplitude of the input field.
Because of this nonlinearity, the readout resonator responds differently to noise than to a readout pulse.
This mechanism contrasts with a conventional Purcell filter, which utilizes the frequency difference between the resonator and the qubit to selectively suppress the energy relaxation of the qubit.

We first show in Sec.\,\ref{sec2} that the conventional dispersive readout with a linear Purcell filter suffers from a fundamental trade-off relation between the measurement rate and the photon-noise-induced dephasing rate.
Then, we describe in Sec.\,\ref{sec3} how a nonlinear Purcell filter enables us to overcome this trade-off by automatically increasing the sensitivity of the qubit to the input field with the application of a readout pulse.
The idea of a self-deactivating filter originates from the ``Josephson quantum filter,'' which is a nonlinear filter that suppresses the coupling of a qubit to a waveguide but saturates and deactivates with the application of a qubit control pulse~\cite{koshino202001protection,kono202007breaking,iakoupov202302saturable}.
To experimentally demonstrate the nonlinear Purcell filter, we fabricate the device described in Sec.\,\ref{ssec:device} and measure the enhancement of the noise tolerance in Sec.\,\ref{ssec:noise-tolerance}.
In Sec.\,\ref{ssec:readout}, we demonstrate fast, high-fidelity readouts, taking advantage of the bifurcation of the nonlinear filter to enhance the separation of the readout signal.
This technique is similar in spirit to the readout schemes that utilize the bifurcation of a nonlinear readout resonator~\cite{siddiqi200602dispersive,lupascu200603highcontrast,mallet200909singleshot,vijay200911invited,ong201104circuit,boissonneault201208improved,dassonneville202210qubit}.
In Sec.\,\ref{ssec:gate-operations}, we demonstrate that the nonlinear Purcell filter is effective even during gate operations.

\section{LIMITATION OF THE CONVENTIONAL SCHEME} \label{sec2}

Without a nonlinear filter, optimizing a readout resonator for a fast, high-fidelity dispersive readout inevitably increases the sensitivity of the qubit to the residual noise photons in the resonator.
This trade-off relation can be quantified as follows.
The photon-noise-induced dephasing rate is proportional to the average noise-photon number $\bar{n}_\mr{noise}$ for the case of weak thermal noise ($\bar{n}_\mr{noise} \ll 1$) and is given by~\cite{clerk200704using,yan201611flux}
\begin{equation} \label{eq:Gamma-th}
    \Gamma_\phi^\mr{noise} =
    \frac{\kappa_\mr{eff} \chi_{qc}^2}{\kappa_\mr{eff}^2 + \chi_{qc}^2}
    \bar{n}_\mr{noise}.
\end{equation}
Here, $\kappa_\mr{eff}$ is the linewidth of the resonator (the effective linewidth if a bandpass Purcell filter~\cite{jeffrey201405fast,sete201507quantum} is used), and $\chi_{qc}$ is the full dispersive shift.
Note that, since the resonator is highly overcoupled to a waveguide, the noise-photon number depends only on the noise spectral density in the waveguide and not on the resonator linewidth or the temperature of the intrinsic bath.
In comparison, the measurement-induced dephasing rate~\cite{schuster200503ac,gambetta200610qubitphoton,blais202105circuit}, which equals the maximum measurement rate~\cite{clerk201004introduction}, is given by
\begin{equation} \label{eq:Gamma-meas}
    \Gamma_\phi^\mr{meas} =
    \frac{2 \kappa_\mr{eff} \chi_{qc}^2}{\kappa_\mr{eff}^2 + \chi_{qc}^2}
    \bar{n}_\mr{meas}.
\end{equation}
Here, $\bar{n}_\mr{meas}$ is the average photon number in the resonator, which is assumed to have reached the steady state under the application of a readout tone.
The frequency of the readout tone is assumed to be the average of the qubit-dressed resonator frequencies corresponding to the two qubit states.
Equations~\eqref{eq:Gamma-th} and~\eqref{eq:Gamma-meas} are similar except for the factor of 2 that originates from the difference between the photon-noise spectra of a thermal state and a coherent state~\cite{yan201806distinguishing}.
The similarity is fundamental because they both describe a process where the information about the qubit state is carried away by the photons decaying out of the resonator.
Therefore, increasing the quantity $\kappa_\mr{eff} \chi_{qc}^2 / (\kappa_\mr{eff}^2 + \chi_{qc}^2)$ not only helps improve the speed and fidelity of the readout but inevitably makes the device more sensitive to the noise photons in the readout resonator.

The ``dynamic range of dephasing rate'' $\Gamma_\phi^\mr{meas} / \Gamma_\phi^\mr{noise}$ quantifies how quickly the qubit can be measured relative to the dephasing rate during its idle time and gate operations.
This is an important figure of merit for applications such as quantum error correction, where ancillary qubits containing the error syndromes need to be measured quickly while their dephasing-induced gate errors need to be minimized.
Since the dephasing rates are proportional to the photon numbers, the dynamic range of the dephasing rates is determined by that of the photon numbers:
\begin{equation}
    \frac{\Gamma_\phi^\mr{meas}}{\Gamma_\phi^\mr{noise}}
    = 2 \frac{\bar{n}_\mr{meas}}{\bar{n}_\mr{noise}}.
\end{equation}
A readout-tone photon number as high as $\bar{n}_\mr{meas} \sim 10^3$~\cite{hassani202307inductively} and a noise-photon number as low as $\bar{n}_\mr{noise} \sim 10^{-4}$~\cite{wang201901cavity} have been reported, suggesting that a dynamic range of up to $\Gamma_\phi^\mr{meas} / \Gamma_\phi^\mr{noise} \sim 10^7$ should be feasible with current technology.
However, in general, a large $\bar{n}_\mr{meas}$ degrades the quantum-nondemolition~(QND) properties of the readout via effects such as measurement-induced energy relaxation~\cite{boissonneault200806nonlinear,picot200810role,boissonneault200901dispersive,serban201002relaxation,slichter201210measurementinduced,harrington201706quantum,malekakhlagh202004lifetime,petrescu202004lifetime,thorbeck202305readoutinduced} and leakage into higher energy levels of the qubit~\cite{sank201611measurementinduced,gusenkova202106quantum,malekakhlagh202202optimization,shillito202209dynamics,khezri202212measurementinduced,cohen202304reminiscence}.
On the other hand, reducing $\bar{n}_\mr{noise}$ requires cooling the microwave components in the readout waveguide, which is becoming increasingly difficult because the electron--phonon coupling diminishes at low temperatures~\cite{yeh201706microwave,wang201901cavity}.
Therefore, increasing $\bar{n}_\mr{meas} / \bar{n}_\mr{noise}$ is not a sustainable approach to improving the dephasing time $T_2$ relative to the time required to perform a readout.
This calls for the introduction of nonlinearity that breaks the proportionality between the dephasing rates and the photon numbers.

\section{NONLINEAR PURCELL FILTER} \label{sec3}

\begin{figure}
    \centering
    \includegraphics{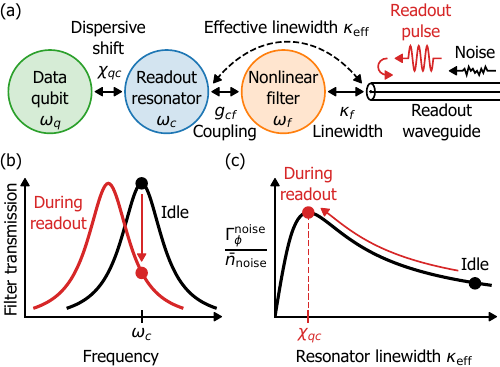}
    \caption{Suppression of photon-noise-induced qubit dephasing with a nonlinear Purcell filter.
    (a)~Schematic of the system. A nonlinear filter resonator is inserted between the readout resonator and the readout waveguide.
    It acts as a tunable coupler that automatically reduces the effective linewidth $\kappa_\mr{eff}$ of the readout resonator with the application of a readout pulse.
    (b)~Shift of the filter passband induced by the application of a readout pulse.
    This decreases the transmission of the filter and therefore the effective linewidth $\kappa_\mr{eff}$ of the readout resonator.
    Here, we have ignored the effect of the bifurcation of the nonlinear filter.
    (c)~Ratio between the photon-noise-induced dephasing rate $\Gamma_\phi^\mr{noise}$ and the average noise-photon number $\bar{n}_\mr{noise}$ in the resonator as a function of the resonator linewidth $\kappa_\mr{eff}$.
    The nonlinear filter is designed such that $\kappa_\mr{eff} \gg \chi_{qc}$~($\kappa_\mr{eff} = \chi_{qc}$) without~(with) the application of a readout pulse, where $\chi_{qc}$ is the full dispersive shift of the readout resonator due to the data qubit.}
    \label{fig1}
\end{figure}

Our nonlinear Purcell filter modifies the relation between the qubit dephasing rate and the photon number in the readout resonator by automatically changing the effective linewidth of the resonator.
Figure~\ref{fig1} shows an intuitive interpretation of how the filter works.
The filter is a Kerr nonlinear resonator inserted between the readout resonator and the readout waveguide.
It can take the form of a weakly anharmonic superconducting qubit such as the transmon qubit~\cite{koch200710chargeinsensitive}, which has the advantage that it can be fabricated along with the data qubit.
The filter is designed to be resonant with the readout resonator so that when there is no readout pulse, the effective linewidth $\kappa_\mr{eff}$ of the resonator is significantly larger than the dispersive shift $\chi_{qc}$.
Then, the photon-noise-induced dephasing of the data qubit is suppressed because, as expressed in Eq.\,\eqref{eq:Gamma-th}, the ratio between the dephasing rate $\Gamma_\phi^\mr{noise}$ and the average noise-photon number $\bar{n}_\mr{noise}$ approaches zero for $\kappa_\mr{eff} \gg \chi_{qc}$.
When the linewidth of the readout resonator is sufficiently large, the noise photons stay in the resonator for only a short time and do not acquire much information about the qubit state.
Therefore, by maintaining the condition $\kappa_\mr{eff} \gg \chi_{qc}$ while the qubit is idle or undergoing gate operations, the nonlinear filter protects the qubit from the photon-noise-induced dephasing.

When a readout pulse is applied, the passband of the filter shifts because of its Kerr nonlinearity, decreasing the effective linewidth $\kappa_\mr{eff}$ of the readout resonator.
The nonlinearity of the filter and the amplitude of the readout pulse can be chosen such that the condition $\kappa_\mr{eff} = \chi_{qc}$ is met during the readout pulse.
This condition maximizes the measurement-induced dephasing rate expressed by Eq.\,\eqref{eq:Gamma-meas} and therefore the measurement rate by the readout pulse.
Thus, the nonlinear filter enables a fast, high-fidelity readout while suppressing the adverse effect of the noise photons on the idle-time coherence and the gate-control fidelities of the qubit.

A distinctive feature of the nonlinear Purcell filter is that its effect is self-deactivating: the sensitivity of the qubit to the input field is automatically increased with the application of a readout pulse.
This contrasts with the scheme demonstrated in Ref.\,\citenum{swiadek202307enhancing}, where the dispersive shift $\chi_{qc}$ is manually increased during a readout by controlling the qubit frequency using a flux pulse.
Our scheme has the advantage that it does not require any control signal other than the readout pulse and is applicable to any type of qubit that can be dispersively read out.

\begin{figure}
    \centering
    \includegraphics{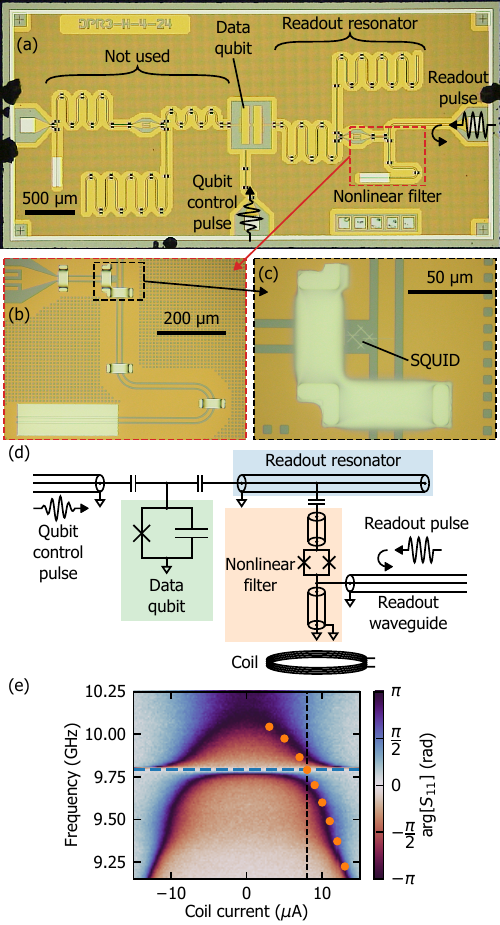}
    \caption{Device containing a nonlinear Purcell filter.
    (a)~Photograph of the entire chip.
    It is made from epitaxially grown titanium nitride~(yellow) and evaporated aluminum~(white) on a silicon substrate~(gray).
    The left side of the chip contains a readout resonator and a linear Purcell filter, which are not used in this work.
    (b)~Enlarged photograph of the nonlinear filter, which is a $\lambda / 4$ resonator interrupted by a superconducting quantum interference device~(SQUID).
    The length of the resonator has been adjusted by short-circuiting its end with an aluminum pad.
    (c)~Enlarged photograph of the SQUID, which was fabricated at the same time as the Josephson junction in the data qubit.
    The nonlinear filter is galvanically connected to the readout waveguide at the right end of the SQUID.
    (d)~Distributed-element circuit model of the relevant part of the device.
    (e)~Phase spectrum of the measured reflection coefficient $S_{11}$ of the nonlinear filter as a function of the current through the coil in the sample holder.
    The blue dashed line represents the frequency of the readout resonator, the orange circles the frequency of the nonlinear filter, and the black dashed line the optimal coil current.}
    \label{fig2}
\end{figure}

\section{EXPERIMENTAL DEMONSTRATION} \label{sec4}

\subsection{Device} \label{ssec:device}

To experimentally demonstrate the nonlinear Purcell filter, we have fabricated the device shown in Figs.\,\ref{fig2}(a)--(c).
The relevant part of the device consists of a data qubit, a readout resonator, and a nonlinear filter and can be represented by the circuit model shown in Fig.\,\ref{fig2}(d).
The data qubit is a transmon~\cite{koch200710chargeinsensitive} with a resonator-dressed qubit frequency of $\omega_q / 2\pi = 8.4969$~GHz and an energy-relaxation time of $T_1 = 17\pm1$~$\mu$s.
The readout resonator is a $\lambda / 2$ resonator with an intrinsic Purcell filter~\cite{sunada202204fast} and has a ground-state-qubit-dressed frequency of $\omega_c / 2\pi = 9.7927$~GHz and a full dispersive shift of $\chi_{qc} / 2\pi = -11.8$~MHz.
The nonlinear filter is a $\lambda / 4$ resonator interrupted by a superconducting quantum interference device~(SQUID) and has a flux-tunable resonance frequency $\omega_f$, an anharmonicity of $\alpha_f / 2\pi = -0.12$~GHz, a coupling strength of $g_{cf} / 2\pi = 88$~MHz to the readout resonator, and an external linewidth of $\kappa_f / 2\pi = 0.31$~GHz.
The intrinsic Purcell filter is implemented because the bandpass-filtering effect of the nonlinear filter does not sufficiently suppress the energy relaxation of the qubit into the readout waveguide.
The energy-relaxation time would have been limited to 47~ns with no Purcell filter and 5~$\mu$s with just the nonlinear filter (for the formula, see Appendix~\ref{app:bandpass-filtering}).
The master equation and the Hamiltonian of the system in the rotating frame of the qubit and the readout tone are given by
\begin{subequations} \label{eqs:master-equation}
\begin{align}
    \frac{\mr{d}}{\mr{d}t} \hat\rho
        & = -i [\hmc{H} / \hbar, \hat\rho]
        + \frac{1}{T_1} \mc{D}[\ketbra{g}{e}] \hat\rho
        + \kappa_f \mc{D}[\hf] \hat\rho, \\
    \hmc{H} / \hbar
        & = (\delta_c + \chi_{qc} \ketbra{e}{e}) \hcd \hc
        + \delta_f \hfd \hf
        + \frac{\alpha_f}{2} (\hfd)^2 \hf^2 \notag \\
        & \qquad \qquad + g_{cf} (\hcd \hf + \hc \hfd)
        + \frac{\Omega}{2} (\hf + \hfd).
\end{align}
\end{subequations}
Here, $\ket{g}$ and $\ket{e}$ are the ground and excited states of the qubit, $\hc$ and $\hf$ are the annihilation operators of the readout resonator and the filter, $\delta_c \coloneqq \omega_c - \omega$ and $\delta_f \coloneqq \omega_f - \omega$ are their frequencies in the rotating frame, $\omega$ and $\Omega$ are the frequency and amplitude of the readout tone, and $\mc{D}[\hf] \hat\rho \coloneqq \hf \hat\rho \hf^\dag - \frac{1}{2} (\hf^\dag \hf \hat\rho + \hat\rho \hf^\dag \hf)$.

While not necessary, the filter in our device is designed to be tunable by applying a magnetic field through the SQUID.
This is to demonstrate the robustness of the filter to parameter variations in the fabrication and to determine the device parameters using the method detailed in Appendix~\ref{app:determining-parameters}.
Figure~\ref{fig2}(e) is obtained by sweeping the current through the coil that generates the magnetic field and measuring the reflection coefficient of the filter.
The optimal operating point is at the coil current of 8~$\mu$A, where the filter is resonant with the readout resonator.
At this point, they hybridize into two modes with frequencies $\omega_\pm$ and linewidths $\kappa_\pm$ given by
\begin{equation}
    \omega_\pm + i \frac{\kappa_\pm}{2}
    = \omega_c + i \frac{\kappa_f}{4}
    \pm \sqrt{g_{cf}^2 - \left(\frac{\kappa_f}{4}\right)^{\!2}}.
\end{equation}
Because this device satisfies $4 g_{cf} \ge \kappa_f$, the linewidth of the filter is equally distributed to the two modes, i.e., $\kappa_+ = \kappa_- = \kappa_f / 2$.
This ensures that the photon-noise-induced dephasing due to both of the hybridized modes are suppressed, which requires $\kappa_+, \kappa_- \gg \chi_{qc}$.

\subsection{Noise tolerance} \label{ssec:noise-tolerance}

\begin{figure}
    \centering
    \includegraphics{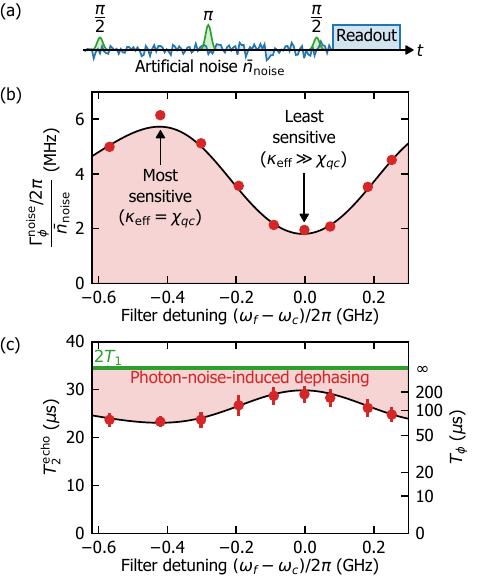}
    \caption{Evaluation of the noise tolerance.
    (a)~Pulse sequence.
    We measure the Hahn-echo dephasing time $T_2^\mr{echo}$ of the qubit while injecting artificial noise through the readout waveguide.
    (b)~Measured noise sensitivity in terms of the photon-noise-induced dephasing rate $\Gamma_\phi^\mr{noise}$ per noise-photon number $\bar{n}_\mr{noise}$~(red circles).
    The theory curve~(black line) is calculated using the device parameters measured independently.
    (c)~Dephasing time $T_2^\mr{echo}$ measured without the artificial noise~(red circles).
    The corresponding pure dephasing time indicated on the right axis is calculated as $T_\phi = [(T_2^\mr{echo})^{-1} - (2 T_1)^{-1}]^{-1}$, where $T_1$ is the energy-relaxation time.
    The error bars represent the standard deviation of the measurements repeated over three hours.
    The theory curve~(black line) assumes a residual noise-photon number of $\bar{n}_\mr{noise} = 4 \times 10^{-4}$ in the fundamental mode of the readout resonator and that there is no other source of pure dephasing.}
    \label{fig3}
\end{figure}

We evaluate the noise tolerance of the qubit in our device by measuring the Hahn-echo dephasing time $T_2^\mr{echo}$ while injecting artificial noise through the readout waveguide.
Figure~\ref{fig3}(a) shows the pulse sequence for the experiment.
White noise in the frequency range of 9.425--10.425~GHz is generated using the arbitrary-waveform generator and the microwave mixer that are used to generate the readout pulse.
This noise emulates the spectral component of the environmental noise that is resonant with the readout resonator and the filter.
Using the band-limited white noise with a power spectral density of $\hbar \omega_c \bar{n}_\mr{noise}$, we can generate $\bar{n}_\mr{noise}$ noise photons in the readout resonator (the power at the device is calibrated in Appendix~\ref{app:determining-parameters}).
Figure~\ref{fig3}(b) shows the measured noise sensitivities in terms of the photon-noise-induced dephasing rate $\Gamma_\phi^\mr{noise}$ per noise-photon number $\bar{n}_\mr{noise}$.
The data are consistent with the theory curve calculated using the device parameters (for the formula, see Appendix~\ref{app:photon-noise-induced-dephasing}).
The noise sensitivity takes the minimum when the filter is resonant with the readout resonator.
The filter is effective even when it is detuned by approximately 100~MHz, which means that precise tunability is not required and the SQUID can be replaced by a Josephson junction in the future.
Conversely, the noise sensitivity takes the maximum when the filter is detuned from the readout resonator by $-$0.42~GHz.
The detuning decreases the effective linewidth $\kappa_\mr{eff}$ of the readout resonator, meeting the condition $\kappa_\mr{eff} = \chi_{qc}$ that maximizes the photon-noise-induced dephasing.
Because the measurement-induced dephasing is also maximized, this operating point would be optimal for a fast, high-fidelity readout if the filter were linear.
Taking the ratio between the maximum and minimum noise sensitivities, we find that the nonlinear filter in our device enhances the noise tolerance by a factor of 3.

Interestingly, the ability to tune the noise sensitivity enables us to estimate the level of noise in the experimental setup.
Figure~\ref{fig3}(c) shows $T_2^\mr{echo}$ measured without the artificial noise.
The corresponding pure dephasing time is calculated as $T_\phi = [(T_2^\mr{echo})^{-1} - (2 T_1)^{-1}]^{-1}$, where $T_1$ is the energy-relaxation time.
Note that $T_1$ has negligible dependence on the filter detuning because it is overwhelmingly determined by the intrinsic energy relaxation of the qubit.
From the dependence of the pure dephasing rate on the filter detuning, we estimate the average noise-photon number to be $\bar{n}_\mr{noise} = 4 \times 10^{-4}$.
This is on par with the best reported values~\cite{yeh201706microwave,yan201806distinguishing,wang201901cavity,somoroff202306millisecond}, in part due to the relatively high frequency of our readout resonator.
The small $\bar{n}_\mr{noise}$ and the relatively short intrinsic $T_1$ of the qubit make the enhancement of noise tolerance less pronounced in Fig.\,\ref{fig3}(c) than in Fig.\,\ref{fig3}(b).
Note that the theory curve in Fig.\,\ref{fig3}(c) assumes that there is no source of pure dephasing other than the residual noise photons in the fundamental mode of the readout resonator.
The theory curve lies within the error bars of the measured $T_2^\mr{echo}$, which means that the other sources of dephasing, including the residual noise photons in the higher harmonics of the readout resonator, are negligible within the uncertainties of our measurements.
Using a qubit with a longer intrinsic $T_1$ will enable us to evaluate the other dephasing processes as well as to observe a threefold enhancement of $T_2^\mr{echo}$.

\subsection{Readout} \label{ssec:readout}

\begin{figure}
    \centering
    \includegraphics{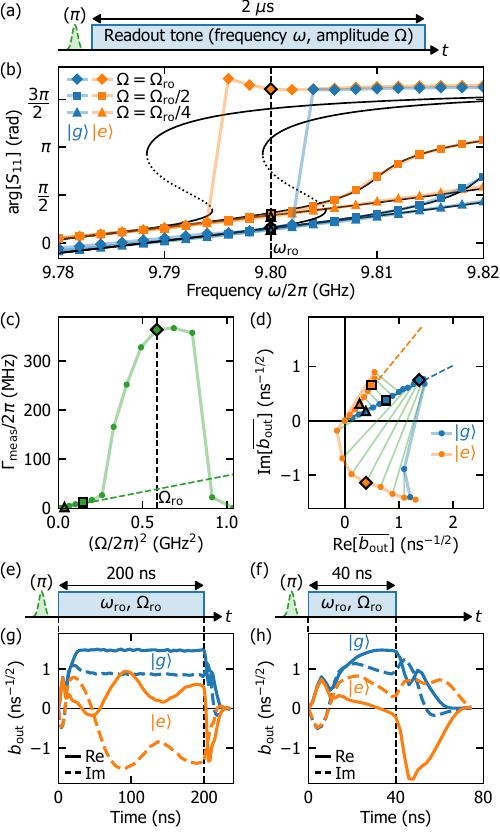}
    \caption{Dispersive readout through a nonlinear Purcell filter.
    (a)~Pulse sequence for measuring the reflection coefficients of the nonlinear filter with the qubit in $\ket{g}$ and $\ket{e}$.
    (b)~Phase spectra of the measured reflection coefficients $S_{11}$ at various amplitudes $\Omega$.
    $\omega_\mr{ro}$ and $\Omega_\mr{ro}$ denote the frequency and amplitude of the readout pulses used in the single-shot readout experiments.
    The fitted curves~(black lines) are calculated using the classical model described in the main text.
    A marker with a black border has a corresponding data point in the following plots.
    (c)~Measurement rate $\Gamma_\mr{meas}$ as a function of the squared amplitude $\Omega^2$ of the readout tone at $\omega_\mr{ro} / 2 \pi = 9.8$~GHz.
    The measurement rate is calculated as $\Gamma_\mr{meas} = |\overbar{b_\mr{out}^e} - \overbar{b_\mr{out}^g}|^2 / 2$, where $\overbar{b_\mr{out}^{g,e}}$ are the average complex amplitudes of the reflected readout tones with the qubit in $\ket{g}$ and $\ket{e}$, respectively.
    If the filter were linear, the measurement rate would be proportional to the squared amplitude~(dashed line).
    (d)~Measured $\overbar{b_\mr{out}^{g,e}}$, normalized such that $|\overbar{b_\mr{out}^{g,e}}|^2$ equal the photon flux.
    The green lines correspond to the signal separation $|\overbar{b_\mr{out}^e} - \overbar{b_\mr{out}^g}|$ used to calculate the measurement rate.
    (e),~(f)~Pulse sequences for the single-shot readout experiments.
    (g),~(h)~Demodulated and averaged readout signals measured without pumping the Josephson parametric amplifier~(JPA).
    Histograms of the integrated readout signal (with the JPA pump for the 40-ns readout) are shown in Appendix~\ref{app:readout-budget}.}
    \label{fig4}
\end{figure}

Here, we demonstrate that the enhanced noise tolerance does not come at the cost of reduced readout fidelity.
The frequency and amplitude of the readout tone are optimized using the pulse sequence shown in Fig.\,\ref{fig4}(a), which measures the reflection coefficients of the nonlinear filter with the qubit in $\ket{g}$ and $\ket{e}$.
Figure~\ref{fig4}(b) shows the phase spectra of the measured reflection coefficients for three different amplitudes of the readout pulse.
The measurement rate by the readout tone is calculated using the semiclassical approximation $\Gamma_\mr{meas} = |\overbar{b_\mr{out}^e} - \overbar{b_\mr{out}^g}|^2 / 2$, where $\overbar{b_\mr{out}^{g,e}}$ are the average complex amplitudes of the reflected readout tones with the qubit in $\ket{g}$ and $\ket{e}$, respectively~\cite{boissonneault201202backaction,laflamme201209weak,clerk201004introduction}.
Here, the complex amplitudes are normalized such that $|\overbar{b_\mr{out}^{g,e}}|^2$ equal the photon flux.
Figure~\ref{fig4}(c) shows the measurement rate as a function of the squared amplitude of a readout pulse at $\omega_\mr{ro} / 2 \pi = 9.8$~GHz.
Figure~\ref{fig4}(d) shows the measured $\overbar{b_\mr{out}^{g,e}}$ used to calculate the measurement rate.

Using the frequency of $\omega_\mr{ro} / 2 \pi = 9.8$~GHz and the amplitude of $\Omega_\mr{ro} / 2 \pi = 0.76$~GHz chosen to maximize the measurement rate, we demonstrate single-shot readouts with 200-ns and 40-ns readout pulses.
Figures~\ref{fig4}(e) and~(f) show the pulse sequences, and Figs.\,\ref{fig4}(g) and~(h) show the demodulated and averaged readout signals.
For the 200-ns readout, only a high-electron-mobility-transistor~(HEMT) amplifier is used as the cryogenic amplifier.
For the 40-ns readout, we also use a Josephson parametric amplifier~(JPA) before the HEMT amplifier.
Following the experimental protocol detailed in Ref.\,\citenum{sunada202204fast}, a readout fidelity of 98.3\% and a QND fidelity of 97.8\% are obtained for the 200-ns readout, and a readout fidelity of 99.4\% and a QND fidelity of 99.2\% are obtained for the 40-ns readout.
The error budgets for the readouts are analyzed in Appendix~\ref{app:readout-budget}.
After a readout, it only takes 10~ns for the number of photons in the readout resonator to decay from one to $10^{-4}$ because the linewidths of the hybridized resonator--filter modes revert to $(\kappa_f / 2) / 2 \pi = 0.16$~GHz.
This allows the gate operations to resume quickly after a readout without implementing an active resonator reset~\cite{mcclure201601rapid,bultink201609active} or qubit cloaking~\cite{lledo202211cloaking}.

Figure~\ref{fig4}(c) also shows the result expected for a linear filter, where the measurement rate would be proportional to the squared amplitude of the readout tone.
Equalizing the sensitivity to the photon noise, the measurement rate of our device is enhanced by a factor of 9 relative to the linear case.
This exceeds the expectation based on Fig.\,\ref{fig3}(b), where we have shown that the photon-noise-induced and measurement-induced dephasing rates can be enhanced by up to a factor of 3 by changing the effective linewidth of the readout resonator.
The additional enhancement of the measurement rate comes from the bifurcation of the nonlinear filter.
This can be understood by considering the steady-state solutions to the classical equations of motion for the mean fields of the readout resonator $c \coloneqq \expect{\hc}$ and the nonlinear filter $f \coloneqq \expect{\hf}$:
\begin{subequations} \label{eqs:classical-eom}
\begin{align}
    \frac{\mr{d}c}{\mr{d}t} & = -i \delta_c' c - i g_{cf} f, \\
    \frac{\mr{d}f}{\mr{d}t} & =
        \left[ -i (\delta_f + \alpha_f |f|^2) - \frac{\kappa_f}{2} \right] f
        - i g_{cf} c - i \frac{\Omega}{2}.
\end{align}
\end{subequations}
Here, the resonator frequency $\delta_c'$ is either $\delta_c$ or $\delta_c + \chi_{qc}$ depending on the qubit state.
As shown in Fig.\,\ref{fig4}(b), the classical model is a good fit to the spectra measured with small amplitudes and is used to determine the anharmonicity of the filter.
As the amplitude is increased, the classical model bifurcates around a qubit-state-dependent frequency.
The measured readout signal for $\ket{g}$ is on the low-excitation branch of the bifurcation with $|f|^2 \approx 0.4$ and $|c|^2 \approx 20$.
The measured readout signal for $\ket{e}$ deviates significantly from the classical model but is closer to the high-excitation branch with $|f|^2 \approx 4$ and $|c|^2 \approx 60$.
The deviation is likely due to the fact that the average photon number $|c|^2$ in the readout resonator is significantly larger than the critical photon number $n_\mr{crit} = 12$~\cite{blais202105circuit}, which invalidates our models given in Eqs.\,\eqref{eqs:master-equation} and~\eqref{eqs:classical-eom}.
The large photon number in the readout resonator also gives rise to the readout-induced state-flip errors observed in the error budget of the single-shot readouts (Appendix~\ref{app:readout-budget}).
Because the readout signals for $\ket{g}$ and $\ket{e}$ are on different branches of the bifurcation, a large separation $|\overbar{b_\mr{out}^e} - \overbar{b_\mr{out}^g}|$ can be realized.
Thus, the nonlinear Purcell filter can not only suppress the photon-noise-induced qubit dephasing but also enhance the measurement rate beyond what is possible with a linear filter.

\begin{figure}
    \centering
    \includegraphics{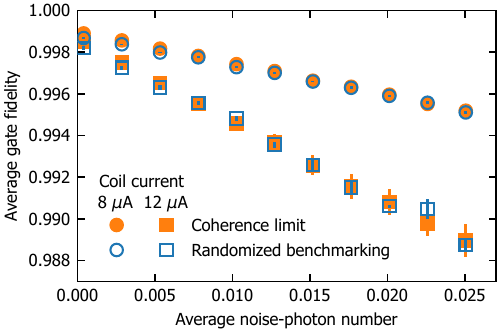}
    \caption{Noise tolerance of the qubit gate fidelity.
    We evaluate the average gate fidelity by randomized benchmarking while injecting artificial noise through the readout waveguide.
    The coherence limit is calculated from $T_1$ and $T_2^\mr{echo}$ measured with the artificial noise being applied.
    The coil current is set to 8~$\mu$A and 12~$\mu$A, where the noise sensitivity takes the minimum and maximum, respectively.}
    \label{fig5}
\end{figure}

\subsection{Gate operations} \label{ssec:gate-operations}

We demonstrate that the nonlinear filter is effective even during gate operations by measuring the average gate fidelity while injecting artificial noise through the readout waveguide.
Randomized benchmarking~(RB)~\cite{magesan201105scalable,magesan201204characterizing} is performed to evaluate the average fidelity of a Clifford gate consisting of two 20-ns qubit control pulses.
The qubit control pulses are recalibrated before each RB experiment to correct for the small ac Stark shift of the qubit due to the artificial noise.
Figure~\ref{fig5} shows the results of the RB experiments.
The coherence limits for the average gate fidelities are calculated from $T_1$ and $T_2^\mr{echo}$ measured with the artificial noise being applied.
We are able to achieve coherence-limited gate fidelities, demonstrating that the $T_2$ enhancement by the nonlinear filter is unaffected by the qubit control pulses.
This is because the readout resonator acts as a bandpass filter and blocks the qubit control pulses from driving the nonlinear filter.
Thus, the nonlinear filter suppresses the photon-noise-induced dephasing not only when the qubit is idle but also during a gate operation.
This is important for applications such as fault-tolerant quantum computation, where the qubits spend most of their time undergoing gate operations.

\section{CONCLUSIONS}

In conclusion, we have proposed and demonstrated a nonlinear Purcell filter that suppresses the dephasing of a superconducting qubit due to noise photons in the readout resonator without sacrificing the readout performance.
We have first described a fundamental limitation of the conventional dispersive readout: $T_2$ relative to the time required to perform a readout is limited by the achievable dynamic range of the photon number in the readout resonator.
We have then shown that the nonlinear Purcell filter enables us to overcome this limitation by automatically tuning the sensitivity of the qubit to the photons in the readout resonator.
We have also shown that the bifurcation of the nonlinear filter can be utilized to further enhance the separation of the readout signal.
The nonlinear Purcell filter is applicable to any type of qubit that can be dispersively read out.
It is also compatible with large-scale integration because it can be robustly fabricated and does not require any additional control signals.

Optimization of the device parameters based on a numerical model is left for a future work.
In this work, the filter anharmonicity of $\alpha_f / 2 \pi = -0.12$~GHz has been chosen to be significantly smaller than the filter linewidth of $\kappa_f / 2 \pi = 0.31$~GHz solely to enable the analysis based on the classical model given in Eqs.\,\eqref{eqs:classical-eom}.
A larger anharmonicity requires a quantum mechanical treatment but may reduce the readout-induced state-flip errors by allowing us to achieve the maximum measurement rate with fewer photons in the readout resonator.
Another interesting direction is to decrease the dispersive shift $\chi_{qc}$, which increases the enhancement factor of the noise tolerance and the critical photon number of the readout resonator.
Without a nonlinear filter, this would require a narrower linewidth of the readout resonator, which would prevent its fast relaxation to the ground state after a readout.
A nonlinear filter mitigates this problem because the effective linewidth of the readout resonator is reduced only during the readout pulse.

%TC:ignore

\begin{acknowledgments}
This work was supported in part by the University of Tokyo Program of Excellence in Photon Science (XPS), the Japan Society for the Promotion of Science (JSPS) Fellowship (Grant No.\ JP22J13650), the Japan Science and Technology Agency (JST) Exploratory Research for Advanced Technology (ERATO) project (Grant No.\ JPMJER1601), the Ministry of Education, Culture, Sports, Science and Technology (MEXT) Quantum Leap Flagship Program (Q-LEAP) (Grant No.\ JPMXS0118068682), and the JSPS Grant-in-Aid for Scientific Research (KAKENHI) (Grant No.\ JP22H04937).
\end{acknowledgments}

\appendix

\section{SAMPLE AND SETUP}

\begin{figure}
\includegraphics{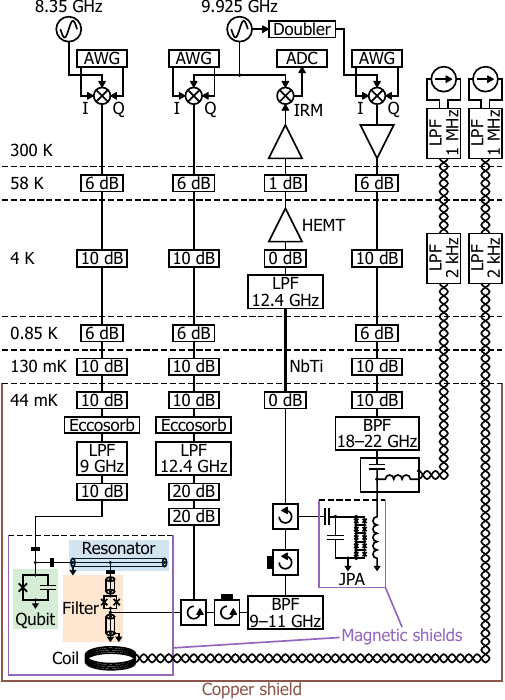}
\caption{Experimental setup.
The unused port on the left side of the chip is connected to a 50-$\Omega$ terminator.
AWG, arbitrary-waveform generator; ADC, analog-to-digital converter; IRM, image reject mixer; LPF, lowpass filter; HEMT, high-electron-mobility transistor; BPF, bandpass filter; JPA, Josephson parametric amplifier.}
\label{fig:setup}
\end{figure}

\begin{table}
\caption{Parameters of the data qubit measured at the coil current of 8~$\mu$A.}
\label{tab:qubit-parameters}
\begin{ruledtabular}
\begin{tabular}{lcr}
Resonator-dressed qubit frequency & $\omega_q / 2 \pi$ & 8.4969~GHz \\
Anharmonicity & $\alpha_q / 2 \pi$ & $-$346.5~MHz \\
Energy-relaxation time & $T_1$ & 17$\pm$1~$\mu$s \\
Ramsey dephasing time & $T_2^*$ & 27$\pm$2~$\mu$s \\
Hahn-echo dephasing time & $T_2^\mr{echo}$ & 29$\pm$2~$\mu$s \\
\end{tabular}
\end{ruledtabular}

\caption{Parameters of the readout resonator. These are assumed to be independent of the coil current.}
\label{tab:resonator-parameters}
\begin{ruledtabular}
\begin{tabular}{lcr}
Ground-state-qubit-dressed frequency & $\omega_c / 2 \pi$ & 9.7927~GHz \\
Qubit--resonator full dispersive shift & $\chi_{qc} / 2 \pi$ & $-$11.8~MHz
\end{tabular}
\end{ruledtabular}

\caption{Parameters of the nonlinear filter resonator. $g_{cf}$ and $\kappa_f$ are assumed to be independent of the coil current.}
\label{tab:filter-parameters}
\begin{ruledtabular}
\begin{tabular}{lcr}
Frequency (coil current $=$ 8 $\mu$A) & $\omega_f / 2 \pi$ & 9.791~GHz \\
Anharmonicity (coil current $=$ 8 $\mu$A) & $\alpha_f / 2 \pi$ & $-$0.12~GHz \\
Coupling strength to readout resonator & $g_{cf} / 2 \pi$ & 88~MHz \\
External linewidth & $\kappa_f / 2 \pi$ & 0.31~GHz \\
\end{tabular}
\end{ruledtabular}
\end{table}

Figure~\ref{fig:setup} shows the setup used to perform the experiments in this work.
Table~\ref{tab:qubit-parameters} shows the parameters of the data qubit measured at the coil current of 8~$\mu$A.
Tables~\ref{tab:resonator-parameters} and~\ref{tab:filter-parameters} show the parameters of the readout resonator and the nonlinear filter, respectively.
The parameters in Tables~\ref{tab:resonator-parameters} and~\ref{tab:filter-parameters}, except for the anharmonicity of the filter, are determined using the method detailed in Appendix~\ref{app:determining-parameters}.
The anharmonicity of the filter is determined using the power dependence of the reflection spectrum shown in Fig.\,\ref{fig4}(b).
These device parameters were measured in the single cooldown in which all the experiments were performed, except for the ones described in Sec.\,\ref{ssec:gate-operations}.
The experiments in Sec.\,\ref{ssec:gate-operations} were performed in a separate cooldown using the same sample and a nominally identical experimental setup.

\section{EXTRACTION OF DEVICE PARAMETERS} \label{app:determining-parameters}

\begin{figure}
    \centering
    \includegraphics{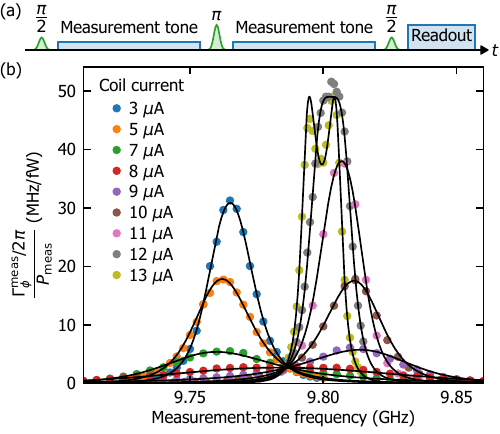}
    \caption{Measurement-induced dephasing rates characterized at various measurement-tone frequencies and coil currents.
    (a)~Pulse sequence for the characterization.
    (b)~Sensitivity to the measurement tone in terms of the measurement-induced dephasing rate $\Gamma_\phi^\mr{meas}$ per measurement-tone power $P_\mr{meas}$~(circles). The black lines are the fits based on Eq.\,\eqref{eq:gamma-meas-fit}.}
    \label{fig:gamma_meas}
\end{figure}

Here, we describe how the device parameters were extracted from the measurement-induced dephasing rates characterized at various measurement-tone frequencies and coil currents.
This method also enables us to calibrate the powers of the measurement tone and artificial noise applied onto the device.
Figure~\ref{fig:gamma_meas} shows the pulse sequence and the result of the experiment.
We obtain the measurement-induced dephasing rates by measuring the Hahn-echo dephasing time $T_2^\mr{echo}$ of the qubit while applying a weak measurement tone.
All the data taken with various coil currents are fitted simultaneously using the model function~\cite{gambetta200610qubitphoton,blais202105circuit}
\begin{equation} \label{eq:gamma-meas-fit}
    \Gamma_\phi^\mr{meas} = \frac{1}{2} |S_{11}^e(\omega_\mr{meas}) - S_{11}^g(\omega_\mr{meas})|^2 \cdot \frac{P_\mr{meas}}{\hbar \omega_\mr{meas}}.
\end{equation}
Here, $\omega_\mr{meas}$ and $P_\mr{meas}$ are the frequency and power of the measurement tone,
\begin{equation} \label{eq:low-power-s11}
    S_{11}^{g,e}(\omega) = 1 - \frac
        {i \kappa_f (\omega - \omega_c')}
        {(\omega - \omega_c') (\omega - \omega_f + i \kappa_f / 2) - g_{cf}^2}
\end{equation}
is the low-power reflection coefficient of the nonlinear filter corresponding to each qubit state, and the resonator frequency $\omega_c'$ is either $\omega_c$ or $\omega_c + \chi_{qc}$ depending on the qubit state.
All the parameters except the filter frequency $\omega_f$ are assumed to be independent of the coil current.
The obtained filter frequencies are plotted in Fig.\,\ref{fig2}(e), and the other parameters obtained are listed in Tables~\ref{tab:resonator-parameters} and~\ref{tab:filter-parameters}.
Note that the anharmonicity of the filter cannot be determined from this experiment because Eq.\,\eqref{eq:low-power-s11} is only valid when the measurement tone is weak enough to avoid any nonlinear effects.

The spectrum of measurement-induced dephasing rates obtained by sweeping the frequency of the measurement tone can be interpreted as the frequency dependence of the sensitivity to the input field.
The peak in the spectrum corresponds to the mode of the readout resonator dressed by the nonlinear filter.
The frequency of the dressed readout mode is lower (higher) when the coil current is lower (higher) because of the repulsion from the filter mode, which has a higher (lower) frequency than the readout resonator.
The width of the peak corresponds to the linewidth of the dressed readout mode, which is equivalent to the effective linewidth $\kappa_\mr{eff}$ of the readout resonator discussed in Fig.\,\ref{fig1} and is largest at the optimal coil current of 8~$\mu$A.
At the coil current of 13~$\mu$A, the peak splits into two because the effective linewidth $\kappa_\mr{eff}$ is smaller than the dispersive shift $\chi_{qc}$.
The area under the curve corresponds to the noise sensitivity $\Gamma_\phi^\mr{noise} / \bar{n}_\mr{noise}$, as the noise spectrum can be considered white within the frequency range shown in the plot.
The height of the peak corresponds to the ratio $\Gamma_\phi^\mr{meas} / \bar{n}_\mr{meas}$ given in Eq.\,\eqref{eq:Gamma-meas}.
As expected, the area and the height are smallest at the optimal coil current of 8~$\mu$A, which means that the qubit is insensitive to noise but cannot be measured efficiently without a nonlinear effect.

\section{SEMICLASSICAL MODEL OF PHOTON-NOISE-INDUCED DEPHASING} \label{app:photon-noise-induced-dephasing}

Here, we present a semiclassical derivation of the photon-noise-induced dephasing rate.
Equation~\ref{eq:Gamma-th} is obtained for the case without a Purcell filter.
Equation~\ref{eq:Gamma-th} is also valid if a conventional bandpass Purcell filter is used, because the hybridization between the filter and the readout resonator is weak.
We also derive the formula for the case where the hybridization is strong.
Because the nonlinear Purcell filter is designed to strongly hybridize with the readout resonator ($\omega_f = \omega_c$ and $4 g_{cf} \ge \kappa_f$), the theory curve in Fig.\,\ref{fig3}(b) is calculated using the latter formula.

Let us consider white noise, which is defined as a random process whose autocorrelation $G_{b_\mr{in} b_\mr{in}}(t)$ is a delta function:
\begin{equation}
    G_{b_\mr{in} b_\mr{in}}(t)
    \coloneqq \lim_{\tau\to\infty} \frac{1}{\tau} \int_{-\tau/2}^{\tau/2} b_\mr{in}(t') b_\mr{in}^*(t' + t) \, \mr{d}t'
    = \bar{n}_\mr{noise} \delta(t).
\end{equation}
Here, $b_\mr{in}(t)$ is the complex amplitude of the classical input field in the readout waveguide.
Using the Wiener--Khinchin theorem, one can see that white noise has a flat power spectral density
\begin{equation}
    S_{b_\mr{in} b_\mr{in}}(\omega)
    = \mc{F}[G_{b_\mr{in} b_\mr{in}}(t)]
    = \bar{n}_\mr{noise},
\end{equation}
where $\mc{F}[\cdot]$ denotes the Fourier transform.
If the input field couples to a resonator with frequency $\omega_c$ and linewidth $\kappa_\mr{eff}$ (the effective linewidth if a bandpass Purcell filter is used), the complex amplitude $c(t)$ of the resonator field is given by
\begin{equation}
    \mc{F}[c(t)] = -\frac{i \sqrt{\kappa_\mr{eff}}}
    {\omega - \omega_c + i \kappa_\mr{eff} / 2}
    \mc{F}[b_\mr{in}(t)].
\end{equation}
Then, the average photon number in the resonator due to the noise can be calculated as
\begin{subequations} \begin{align}
    \overbar{|c(t)|^2}
    & = G_{cc}(0) \\
    & = \mc{F}^{-1}[S_{cc}(\omega)](0) \\
    & = \mc{F}^{-1}\left[
        \left| \frac{\sqrt{\kappa_\mr{eff}}}{\omega - \omega_c + i \kappa_\mr{eff} / 2} \right|^2
        S_{b_\mr{in} b_\mr{in}}(\omega)
    \right](0) \label{eq:noise-photons-3} \\
    & = \bar{n}_\mr{noise} \int_{-\infty}^\infty
        \left| \frac{\sqrt{\kappa_\mr{eff}}}{\omega - \omega_c + i \kappa_\mr{eff} / 2} \right|^2
        \frac{\mr{d}\omega}{2 \pi} \\
    & = \bar{n}_\mr{noise}.
\end{align} \end{subequations}
This shows that the average noise-photon number $\overbar{|c(t)|^2}$ depends only on the noise spectral density of the input field and not on the resonator linewidth.
One can also see from Eq.\,\eqref{eq:noise-photons-3} that only the spectral component of the noise that is resonant with the resonator contributes to the excitation of the resonator.
This is why white noise is a good approximation in this context for other types of noise whose spectra are only locally flat, including thermal noise and the band-limited white noise used in Sec.\,\ref{ssec:noise-tolerance}.

To calculate the photon-noise-induced dephasing rate, we denote the component of the readout signal that contains information about the qubit state as
\begin{equation}
    s(t) \coloneqq b_\mr{out}^e(t) - b_\mr{out}^g(t).
\end{equation}
Here, $b_\mr{out}^{g,e}(t)$ are the output signals in the readout waveguide when the qubit is in $\ket{g}$ and $\ket{e}$, respectively.
Using $s(t)$, the photon-noise-induced dephasing rate can be expressed as~\cite{gambetta200610qubitphoton,blais202105circuit}
\begin{equation}
    \Gamma_\phi^\mr{noise} = \frac{1}{2} \overbar{|s(t)|^2}.
\end{equation}
This can be rewritten as
\begin{subequations} \begin{align}
    \overbar{|s(t)|^2}
    & = G_{ss}(0) \\
    & = \mc{F}^{-1}[S_{ss}(\omega)](0) \\
    & = \mc{F}^{-1}[|S_{11}^e(\omega) - S_{11}^g(\omega)|^2 S_{b_\mr{in} b_\mr{in}}(\omega)](0) \\
    & = \bar{n}_\mr{noise} \int_{-\infty}^\infty |S_{11}^e(\omega) - S_{11}^g(\omega)|^2
        \frac{\mr{d}\omega}{2 \pi},
\end{align} \end{subequations}
where $S_{11}^{g,e}(\omega)$ are the reflection coefficients when the qubit is in $\ket{g}$ and $\ket{e}$, respectively.
Equation~\eqref{eq:Gamma-th} is obtained by substituting the reflection coefficients of a resonator dispersively coupled to a qubit
\begin{equation}
    S_{11}^{g,e}(\omega) = 1 - \frac{i \kappa_\mr{eff}}{\omega - \omega_c' + i \kappa_\mr{eff} / 2},
\end{equation}
where the resonator frequency $\omega_c'$ is either $\omega_c$ or $\omega_c + \chi_{qc}$ depending on the qubit state.
The formula for the theory curve in Fig.\,\ref{fig3}(b) is obtained by substituting the reflection coefficients given in Eq.\,\eqref{eq:low-power-s11}.

\section{PURCELL FILTERING BY THE NONLINEAR FILTER} \label{app:bandpass-filtering}

The nonlinear Purcell filter acts as a bandpass Purcell filter even when it is completely hybridized with the readout resonator.
Here, we show this by calculating the energy-relaxation rate of the qubit through the readout resonator and the nonlinear Purcell filter, ignoring the effect of the intrinsic Purcell filter in the readout resonator.

The quantum Langevin equations for the readout resonator and the nonlinear filter are given by
\begin{subequations}
\begin{align}
    \frac{\mr{d}\hc}{\mr{d}t} & = -i \omega_c \hc - i g_{cf} \hf, \\
    \frac{\mr{d}\hf}{\mr{d}t} & =
        \left( -i \omega_f - \frac{\kappa_f}{2} \right) \hf
        - i g_{cf} \hc - i \sqrt{\kappa_f}\,\hb_\mr{in}(t),
\end{align}
\end{subequations}
where $\hb_\mr{in}(t)$ is the input field in the readout waveguide.
These can be solved for $\hat{c}(t)$ as
\begin{equation}
    \hc(t) = \mc{F}^{-1}\left[
        \frac{g_{cf} \sqrt{\kappa_f}}
        {(\omega - \omega_c)(\omega - \omega_f + i\kappa_f/2) - g_{cf}^2}
    \right] * \hb_\mr{in}(t),
\end{equation}
where $*$ denotes the convolution.
The quantum autocorrelation function of the input field in the vacuum state is given by
\begin{equation}
    G_{\hb_\mr{in} \hb_\mr{in}}(t)
    \coloneqq \expect{\hb_\mr{in}(0) \hbd_\mr{in}(t)}
    = \delta(t),
\end{equation}
and the quantum noise spectral density by
\begin{equation}
    S_{\hb_\mr{in} \hb_\mr{in}}(\omega)
    \coloneqq \mc{F}[G_{\hb_\mr{in} \hb_\mr{in}}(t)]
    = 1.
\end{equation}
Therefore, the quantum noise spectral density of the annihilation operator $\hat{c}(t)$ of the readout resonator is
\begin{equation}
    S_{\hc\hc}(\omega)
    = \left|
        \frac{g_{cf} \sqrt{\kappa_f}}
        {(\omega - \omega_c)(\omega - \omega_f + i\kappa_f/2) - g_{cf}^2}
    \right|^2.
\end{equation}
Using Fermi's golden rule, the energy-relaxation rate of the qubit into the readout waveguide can be calculated as
\begin{subequations}
\begin{align}
    \Gamma_\mr{ex}
    & = g_{qc}^2 S_{\hc\hc}(\omega_q) \\
    & = \frac{g_{qc}^2 g_{cf}^2 \kappa_f}
        {|(\omega_q - \omega_c)(\omega_q - \omega_f + i\kappa_f/2) - g_{cf}^2|^2},
\end{align}
\end{subequations}
where $g_{qc}$ is the coupling strength between the qubit and the readout resonator.
Using the device parameters listed in Tables~\ref{tab:qubit-parameters}--\ref{tab:filter-parameters} and the qubit--resonator coupling strength calculated as
\begin{equation}
    g_{qc} = \sqrt{\frac
        {(\omega_q - \omega_c)(\omega_q + \alpha_q - \omega_c)}
        {2 \alpha_q}
    \chi_{qc}},
\end{equation}
we obtain an external energy-relaxation rate of $\Gamma_\mr{ex} / 2 \pi = 31$~kHz, which translates to a $T_1$ limit of $1 / \Gamma_\mr{ex} = 5$~$\mu$s.
In comparison, if there were no filter and the readout resonator had a linewidth of $\kappa_f / 2$, the external energy-relaxation rate would be
\begin{equation}
    \Gamma_\mr{ex}
    = \frac{g_{qc}^2 \kappa_f / 2}{|\omega_q - \omega_c + i \kappa_f / 4|^2}
    \approx \left( \frac{g_{qc}}{\omega_q - \omega_c} \right)^{\!\!2} \frac{\kappa_f}{2},
\end{equation}
which evaluates to a $T_1$ limit of $1 / \Gamma_\mr{ex} = 47$~ns.
This means that our nonlinear Purcell filter is suppressing the external energy-relaxation rate of the qubit by 2 orders of magnitude.
The external energy relaxation of our qubit is strongly suppressed by the combination of this bandpass-filtering effect of the nonlinear filter and the intrinsic Purcell filter in the readout resonator.

\section{SINGLE-SHOT READOUTS} \label{app:readout-budget}

\begin{figure}
    \centering
    \includegraphics{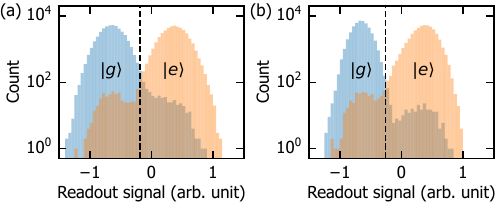}
    \caption{(a),~(b)~Histograms of the integrated readout signals for the 200-ns and 40-ns single-shot readouts obtained using the pulse sequences shown in Figs.\,\ref{fig4}(e) and~(f), respectively.
    The dashed lines represent the thresholds for discriminating the outcome of the readout.}
    \label{fig:readout-hist}
\end{figure}

\begin{figure}
\includegraphics{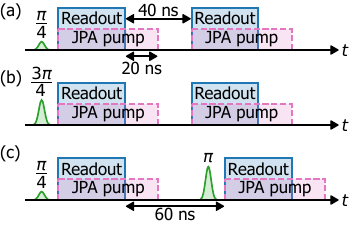}
\caption{Pulse sequences used to analyze the error budget for a readout.
(a) and~(b) are used to obtain the QND fidelity and to determine the separation error. (a) and~(c) are used to obtain the readout fidelity and to determine the upper bound of the state-preparation error.}
\label{fig:readout-budget}
\end{figure}

The waveforms obtained in the single-shot-readout experiments are processed following Ref.\,\citenum{sunada202204fast}.
Figures~\ref{fig:readout-hist}(a) and~(b) show the histograms of the integrated readout signals for the 200-ns and 40-ns single-shot readouts obtained using the pulse sequences shown in Figs.\,\ref{fig4}(e) and~(f), respectively.
The thresholds for discriminating the outcome of the readouts are optimized to maximize the readout fidelities.
The error budgets for these readouts are analyzed using the pulse sequences shown in Fig.\,\ref{fig:readout-budget} and the method detailed in Ref.\,\citenum{sunada202204fast}.
The obtained error budgets are listed in Tables~\ref{tab:f-budget} and~\ref{tab:q-budget}.

\clearpage

\begin{table}
\caption{Error budgets for the readout infidelities of the 200-ns and 40-ns readouts.}
\label{tab:f-budget}
\begin{ruledtabular}
\begin{tabular}{lllrr}
\multicolumn{3}{l}{Readout-pulse duration} & 200~ns & 40~ns \\
\hline
\multicolumn{3}{l}{Readout infidelity} & 1.7\% & 0.6\% \\
\multirow{4}{*}{Error budget $\left\{\rule{0pt}{32pt}\right.$} & \multicolumn{2}{l}{State-preparation error} & $\le$1.2\% & $\le$0.4\% \\
& \multirow{2}{*}{Back action $\left\{\rule{0pt}{14pt}\right.$} & $\ket{e} \to \ket{g}$ & $\le$0.8\% & $\le$0.2\% \\
& & $\ket{g} \to \ket{e}$ & $\le$0.6\% & $\le$0.1\% \\
& \multicolumn{2}{l}{$T_1$ decay} & $\le$0.7\% & $\le$0.2\% \\
& \multicolumn{2}{l}{Separation error} & 0.3\% & 0.2\% \\
\end{tabular}
\end{ruledtabular}
\end{table}

\begin{table}
\caption{Error budgets for the QND infidelities of the 200-ns and 40-ns readouts.}
\label{tab:q-budget}
\begin{ruledtabular}
\begin{tabular}{lllrr}
\multicolumn{3}{l}{Readout-pulse duration} & 200~ns & 40~ns \\
\hline
\multicolumn{3}{l}{QND infidelity} & 2.2\% & 0.8\% \\
\multirow{4}{*}{Error budget $\left\{\rule{0pt}{26pt}\right.$} & \multirow{2}{*}{Back action $\left\{\rule{0pt}{14pt}\right.$} & $\ket{e} \to \ket{g}$ & 0.8\% & 0.2\% \\
& & $\ket{g} \to \ket{e}$ & 0.4\% & 0.1\% \\
& \multicolumn{2}{l}{$T_1$ decay} & 0.7\% & 0.2\% \\
& \multicolumn{2}{l}{Separation error} & 0.4\% & 0.2\% \\
\end{tabular}
\end{ruledtabular}
\end{table}

\FloatBarrier
\nocite{*}  % check for uncited bibliography entries
\bibliography{bibliography}  % use BibTeX, not BibLaTeX

%TC:endignore

\end{document}